\documentclass[pre,amssymb,superscriptaddress,showpacs,floatfix,a4paper,twocolumn]{revtex4-1}

\usepackage{graphicx}
\usepackage{dcolumn}
\usepackage{bm}
\usepackage{color}
\usepackage{amsmath}
\usepackage{amssymb}
\usepackage{xcolor,cancel}

\begin{document}

\title{Magnetic hysteresis behavior of granular manganite La$_{0.67}$Ca$_{0.33}$MnO$_3$ nanotubes}

\author{M. I. Dolz}
\affiliation{Departamento de F\'isica,  
Universidad Nacional de San Luis, 
Instituto de F\'isica Aplicada (INFAP), 
Consejo Nacional de Investigaciones Cient\'ificas y T\'ecnicas (CONICET), 
Chacabuco 917, D5700BWS San Luis, Argentina}
\author{S. D. Calder\'on Rivero}
\affiliation{Departamento de F\'isica,  
Universidad Nacional de San Luis, 
Instituto de F\'isica Aplicada (INFAP), 
Consejo Nacional de Investigaciones Cient\'ificas y T\'ecnicas (CONICET), 
Chacabuco 917, D5700BWS San Luis, Argentina}
\author{H. Pastoriza}
\affiliation{Centro At\'omico Bariloche,
Comisi\'on Nacional de Energ\'ia At\'omica (CNEA),
Consejo Nacional de Investigaciones Cient\'ificas y T\'ecnicas (CONICET),
Av. E. Bustillo 9500, R8402AGP San Carlos de Bariloche, R\'io Negro, Argentina}
\author{F. Rom\'a}
\affiliation{Departamento de F\'isica,  
Universidad Nacional de San Luis, 
Instituto de F\'isica Aplicada (INFAP), 
Consejo Nacional de Investigaciones Cient\'ificas y T\'ecnicas (CONICET), 
Chacabuco 917, D5700BWS San Luis, Argentina}

\begin{abstract}

A silicon micromechanical torsional oscillator is used to measure the hysteresis
loops of two manganite La$_{0.67}$Ca$_{0.33}$MnO$_3$ nanotubes at different temperatures,
applying an external field along its main axes.
These structures are composed of nanograins with a ferromagnetic 
core surrounded by a dead layer. 
Micromagnetic calculations based on the stochastic Landau-Lifshitz-Gilbert equation,
are performed to validate a simple model that allows for quantitatively describing 
the ferromagnetic behavior of the system.
Further simulations are used to analyze the experimental data more in depth
and to calculate the coercive field, the saturation and remanent magnetizations, 
and the effective magnetic volume for single nanotubes, over a wide temperature range.

\end{abstract}

\maketitle

\section{Introduction}

Micrometric and nanometric structures built from a wide variety of materials,
play a preponderant role in the current development of science and technology \cite{Poole2003}.
For example, due to their singular electronic transport and magnetic properties,
low-dimensional perovskite manganite oxide nanostructures have 
promising applications in mesoscopic physics and nanoscale devices \cite{Fert1999,Handoko2010,Li2016}.
At present, these materials continue to be studied experimentally, numerically, and theoretically 
to better understand how their outstanding physical characteristics arise.

Such manganite nanostructures are synthesized by different physical and chemical methods \cite{Li2016}.
In particular, a versatile and inexpensive chemical technique
uses sacrificial porous polycarbonate substrates as templates
to produce nanowires and nanotubes with a disordered ``granular'' structure, i.e.,
formed by an irregular assembly of magnetic nanograins or nanoparticles \cite{Levy2003,Leyva2004,Curiale2004}.
At low temperatures, these manganite nanotubes show a homogeneous ferromagnetic behavior and, 
since the nanograins that compose them are very small, 
basic theoretical approximations indicate that, 
at least at zero temperature, probably, these nanoparticles 
behave like single magnetic domains \cite{Curiale2007}.

Unlike other types of ferromagnetic nanotubes with a homogeneous structure \cite{Wyss2017},
these granular systems have an additional characteristic that makes them very interesting.
Manganite nanoparticles are composed by a ferromagnetic core 
surrounded by a magnetic dead layer \cite{Kaneyoshi1990}.
The existence of this outer shell avoids exchange interactions 
among magnetic moments of contiguous nanoparticles and therefore 
the dominant interaction is the dipolar one \cite{Curiale2007b}.
In other words, these systems constitute the best experimental candidates 
to test the theoretical and simulation findings for models of
low-dimensional disordered arrays of magnetic moments coupled 
via dipolar long-range interactions.      

Manganite nanotubes and nanowires have been studied through different methods. 
Mainly using different experimental techniques, 
the magnetic properties of these nanostructures 
have been determined from measurements of powder samples \cite{Curiale2004,Curiale2007,Curiale2009,Curiale2007b}.
In most cases, these samples are constituted of 
compact packages of randomly oriented nanotubes and,
therefore, the results obtained reflect different aspects of this system:
the orientation magnetic characteristics of the nanotubes  
and the (dipolar) interactions between these nanostructures.

However, in Ref.~\cite{Dolz2008}, using a silicon 
micromechanical torsional oscillator working in its resonant mode,
it was possible to study two single La$_{0.67}$Ca$_{0.33}$MnO$_3$ (LCMO) nanotubes. 
The entire hysteresis loop was measured at a very low temperature 
by applying an external magnetic field along the main axes of the nanotubes.
As was expected, this curve shows a more abrupt behavior than the one observed
for powder samples, i.e., the remanent magnetization for single nanotubes
is greater than for a set of randomly oriented nanostructures,
but the coercive field values for both systems are very close to each other.
In addition, by extrapolating to high external fields,
the saturation magnetization for single nanotubes is obtained,
showing a linear decay with temperature that is very different 
from the one measured for bulk and powder samples. 

These nanostructured materials have been little studied numerically.  
By using Monte Carlo simulations \cite{Cuchillo2008}, 
the hysteresis loops of a model of granular nanotube were calculated.
The authors show that these curves agree qualitatively well with
the experimental data for La$_{0.67}$Sr$_{0.33}$MnO$_3$ manganite nanotubes,
and conclude that the simulations neglecting 
dipole-dipole interaction never adjust to the experiment.

More recently, micromagnetic calculations based on 
the stochastic Landau-Lifshitz-Gilbert (sLLG) equation \cite{Brown1963}
were performed to simulate the real dynamic behavior of a one-dimensional model 
for granular nanotubes \cite{Longone2018}.
Although the shape of the hysteresis loops are not sensitive 
to the choice of the volume of nanograins,
it was observed that this depends largely on the distribution 
of the anisotropy constant of each nanograin. 
Assuming that this quantity is uniformly distributed,
the simulations allowed us to describe reasonably well the
experimental data reported in the literature for single LCMO nanotubes 
and powder samples of this material, measured at $14$ K.

In the present paper we have carried out a more thorough 
and systematic experimental and numerical study 
of these nanostructured magnetic systems.
Following the lines of Ref.~\cite{Dolz2008}, 
we use a silicon micromechanical torsional oscillator 
to perform measurements of LCMO single nanotubes,
extending the measurements to a very wide range of temperatures.
Furthermore micromagnetic simulations are employed first to determine 
which model is most suitable to describe the dynamic of these systems,
and then to analyze the experimental data more in depth.
  
The outline of the paper is as follows.  In Sec.~\ref{Experiments},
we describe how a micromechanical torsional oscillator is used
to measure the hysteresis loops of two LCMO nanotubes,
and also we present our main experimental results. 
Then in Sec.~\ref{Simulations}, we introduce the numerical 
micromagnetic scheme of calculation employed in this paper and,
after carrying out a systemic study, we determine which is the most suitable 
model to describe the experimental data. 
Finally, Sec.~\ref{ResCon} is devoted to describe 
the results and conclusions obtained in this work.      

\section{Experiments \label{Experiments}}

A chemical method in which porous sacrificial substrates 
of polycarbonate are used as templates is employed to synthesize 
a powder sample of manganite LCMO nanotubes \cite{Curiale2007}.
Typically, this technique produces irregular and disordered nanostructures 
that have lengths of about $6 \sim 10$ $\mu$m, external diameter of $700 \sim 800$ nm,
nominal wall thickness of $\sim 60$ nm, and are composed of nanoparticles with a 
characteristic diameter range between $10$ nm and $40$ nm.
The LCMO compound in bulk develop ferromagnetic properties below a
critical temperature of $T_c \cong 273$ K \cite{Cheong2004}. 

\begin{figure}[t!]
\begin{center}
\includegraphics[width=6.5cm,clip=true]{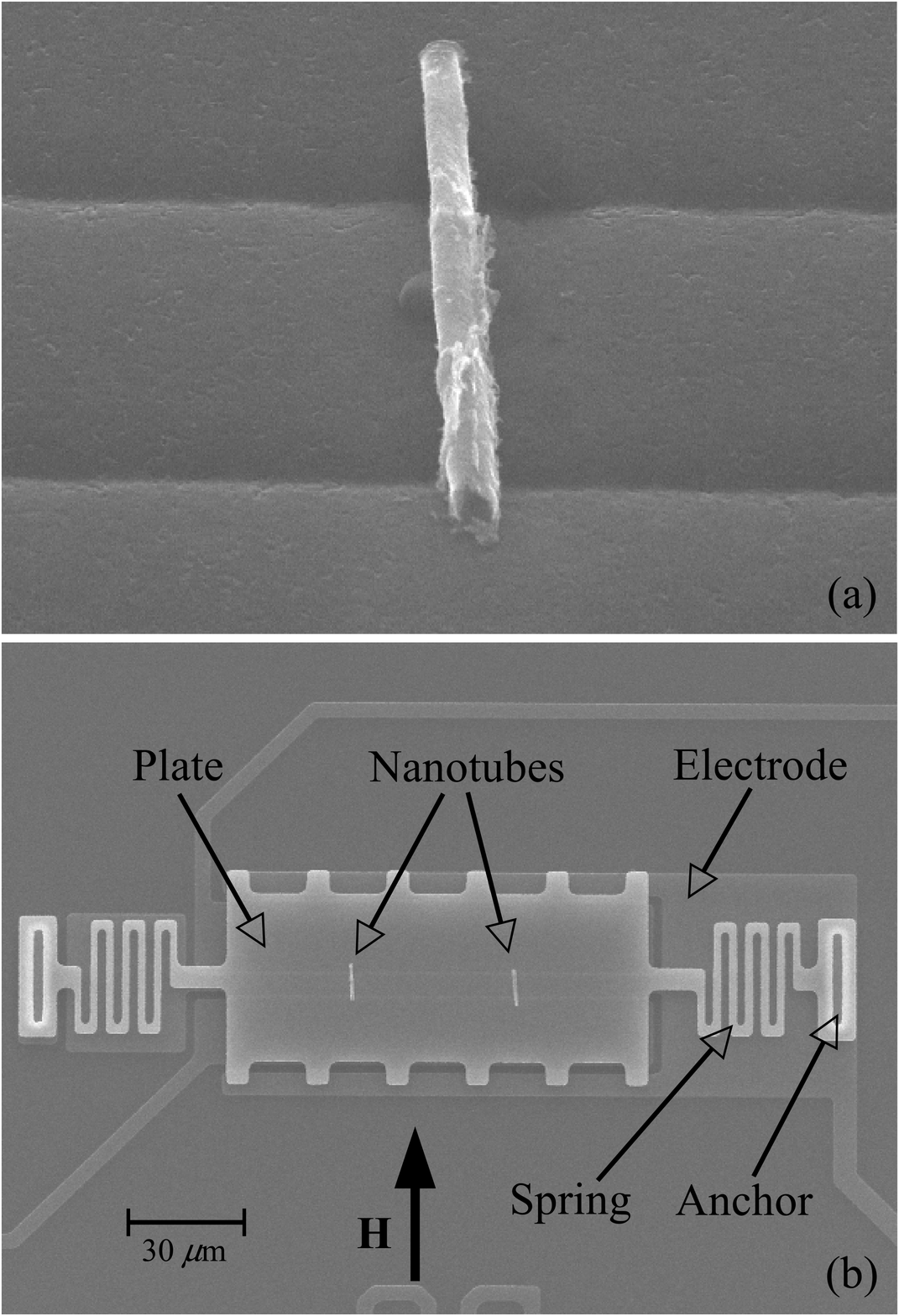}
\caption{SEM images of (a) a LCMO nanotube, and (b) the micromechanical torsional oscillator 
with the two nanotubes stuck on its plate.  The main structural parts of the device
are indicated. } 
\label{figure1}
\end{center}
\end{figure}

Under an optical microscope, a hydraulic micromanipulator was used to handle single nanostructures.
Using a submicrometer drop of Apiezon N grease, 
two LCMO nanotubes were glued on top of a silicon micromechanical torsional oscillator. 
Figure~\ref{figure1}(a) shows a scanning electron microscope (SEM) image 
of one of these nanostructures which, in particular,
has a length of approximately $9.5$ $\mu$m.
The microdevices were manufactured in the MEMSCAP Inc. foundry \cite{MEMSCAP}
and have been previously used as micromagnetometers of high sensitivity 
to study these LCMO nanotubes \cite{Dolz2008,Antonio2010} as well as
mesoscopic samples of a high-$T_c$ superconductor \cite{Dolz2007,Dolz2010}.    
In Fig.~\ref{figure1}(b) we show a top SEM image of the two nanotubes 
placed on the plate of the oscillator.
The magnetic nanostructures are separated by a distance of approximately
40 $\mu$m and are oriented perpendicularly to the rotation axis of the device.
The whole system was cooled under vacuum inside a helium closed-cycle cryogenerator
and a uniform magnetic field, provided by electromagnet, 
was applied along the easy axis of the nanotubes [see Fig.~\ref{figure1}(b)].  
The micromechanical oscillator was actuated electrostatically by means of
a function generator and its movement was sensed capacitively using a lock-in amplifier. 
Additional details of the experimental setup 
can be found in Refs.~\cite{Dolz2008,Antonio2010}. 

We describe now how the microdevice is used to measure 
the magnetic hysteresis behavior of an anisotropic mesoscopic sample.
The resonant frequency of the torsional oscillator with the two nanotubes stuck 
on its plate is
\begin{equation}  
\nu_0=\frac{1}{2\pi}\sqrt{\frac{k_e}{I}}, 
\label{nu_0}
\end{equation}
where $k_e$ is the elastic restorative constant of the serpentine springs and
$I=3.8 \times 10^{-21}$ kg m$^2$ is the moment of inertia of the system
along its center rotational axis.  
Our microdevice has a resonant frequency close to $72.2$ kHz and a quality
factor $Q$ greater than $5\times10^4$, which means that the width of
the resonant peak is less than $2$ Hz.

Because the LCMO nanotubes are ferromagnetic and have a high shape anisotropy 
well below the Curie temperature of the material, 
an external magnetic field $\mathbf{H}$ applied parallel to the plate plane
exerts an additional restoring torque and therefore the new resonant frequency will be      
\begin{equation}  
\nu_r=\frac{1}{2\pi}\sqrt{\frac{k_e+k_M}{I}}. 
\label{nu_r}
\end{equation}
Here, $k_M$ is the effective elastic constant originated by the interaction 
between the magnetization of the sample (of both nanotubes), $\mathbf{M}$, 
and the field $\mathbf{H}$.

\begin{figure}[t!]
\begin{center}
\includegraphics[width=6.5cm,clip=true]{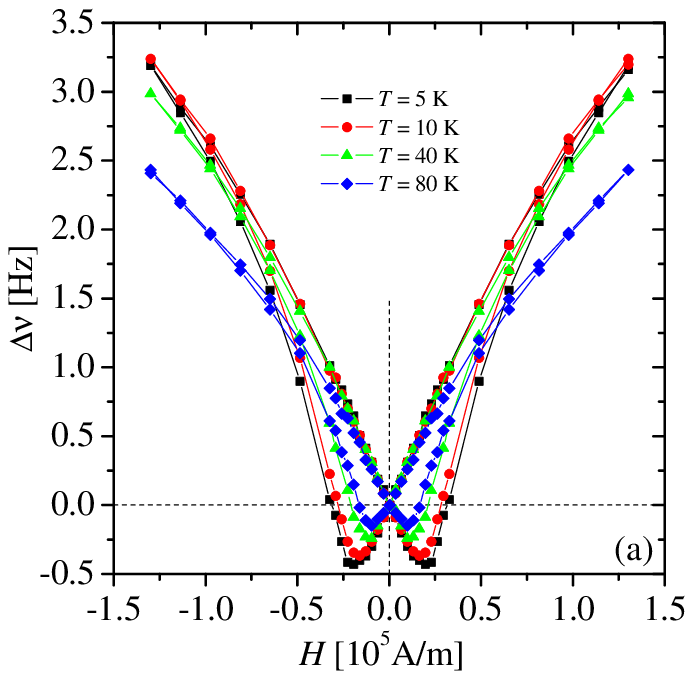}
\includegraphics[width=6.5cm,clip=true]{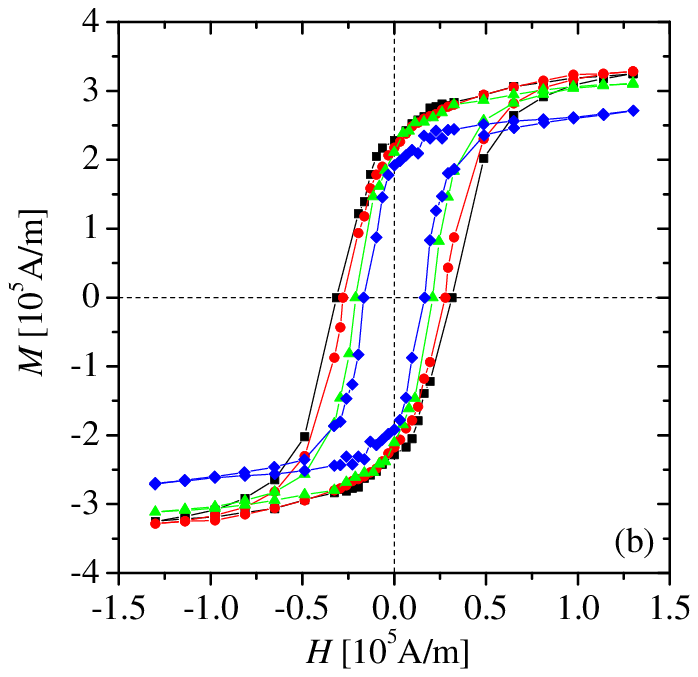}
\caption{(Color online) (a) The changes in the resonant frequency as a function 
of the external field for four different temperatures as indicated. 
(b) The corresponding hysteresis loops calculated using Eq.~(\ref{rootMplus}).} 
\label{figure2}
\end{center}
\end{figure}

The changes in the resonant frequency, $\Delta \nu=\nu_r-\nu_o$,
produced by the external field are presented in Fig.~\ref{figure2} (a). 
Although measurements have been made for temperatures 
$T=5$, $10$, $15$, $20$, $40$, $60$, and $80$ K,
for simplicity we only show curves for four of them. 
Due to the symmetry of the problem, the magnetization 
always points along the direction of easy axes of the nanotubes.
Both the effective elastic constant $k_M$ and $\Delta \nu$, 
are positive (negative) when $\mathbf{M}$ is parallel (antiparallel) to $\mathbf{H}$ \cite{Dolz2007}. 
The curves cross the abscissa axis ($\Delta \nu=0$) 
when the external field is reduced to zero and the sample reaches a remanent magnetization 
or when a reverse field (the coercive one) is applied that cancels the magnetization.    

To calculate the hysteresis loops, we proceed as follows.
Under typical experimental conditions, $\Delta \nu \ll \nu_0$.
Then, from Eqs.(\ref{nu_0}) and  (\ref{nu_r}) it is possible to write that
\begin{equation} 
k_M \simeq 8 \pi^2 I \nu_0 \Delta \nu .  
\end{equation}
This effective elastic constant depends on the magnetic properties of the sample
and can be calculated from energy considerations \cite{Dolz2008,Zijlstra1961,Morillo1998}.
Given the high aspect ratio of the nanotubes (more than 10),
the energy per unit volume of the sample can be written as 
\begin{equation}
U= -\mu_0 \ \mathbf{H} \cdot \mathbf{M} - \frac{K_n}{M^2} (\mathbf{M} \cdot \mathbf{n})^2,    
\end{equation}
where the first term represents the Zeeman interaction and the second the uniaxial anisotropy energy.
$\mu_0$ is the vacuum permeability constant, 
$\mathbf{n}$ is an unit vector pointing along the major (easy) axis of each nanotube 
(parallel to the plate of the micro-oscillator),
and $K_n$ is the uniaxial shape anisotropy constant. 
Keeping constants both the temperature and the magnitude of the external field, $H$,
and due to the amplitude of oscillation being very small (around 1 sexagesimal degree at resonance),
it is possible to write  
\begin{equation}
\frac{1}{8 \pi^2 I \nu_0 \Delta  \nu} \simeq \frac{1}{k_M} = \frac{1}{2K_nV_n}+\frac{1}{MV_n \mu_0 H},
\label{mainEq}
\end{equation}
where $V_n \approx 2.32 \times 10^{-18}$ m$^3$ is the volume of the two nanotubes.  
Equation (\ref{mainEq}) is valid if the module and angle of magnetization, $M$, 
does not change appreciably throughout the oscillation cycle,
a condition that is well satisfied given the small amplitude of oscillation.      

Taking $K_n=\mu_0 M^2 /4$, the value of the shape anisotropy constant for an infinitely long rod \cite{Cullity}, 
Eq.~(\ref{mainEq}) can be written as a quadratic equation in $M$,
\begin{equation}
\Bigg(\frac{H}{2} \Bigg) M^2-(Z \Delta \nu) M - (2Z \Delta \nu H) = 0,
\label{EqM}
\end{equation}
where the constant 
\begin{equation}
Z=\frac{8 \pi^2 I \nu_0}{2 V_n \mu_0}.
\end{equation}
The solution of Eq.~(\ref{EqM}) is 
\begin{equation}
M=\frac{Z \Delta \nu \pm \sqrt{(Z \Delta \nu)^2+4Z \Delta \nu H^2}}{H},
\label{rootM}
\end{equation}
which allows us to calculate the hysteresis loops from the 
experimental curves $\Delta \nu$ vs $H$.
To guarantee that $dM/dH \ge 0$, the positive (negative) root 
of Eq.~(\ref{rootM}) should be taken when $\Delta \nu>0$ ($\Delta \nu<0$).
Nevertheless, this rule can be avoided by rewriting Eq.~(\ref{rootM}) as
\begin{equation}
M=\frac{Z \Delta \nu}{H} \Bigg[ 1 + \sqrt{1+\frac{4 H^2}{Z \Delta \nu}}  \Bigg],
\label{rootMplus}
\end{equation}
where only the plus sign has been considered. 

In Fig.~\ref{figure2} (b), we present the hysteresis loops calculated with the previous procedure.
These curves show a typical temperature behavior: When $T$ increases 
the coercive field, $H_c$, the remanent magnetization, $M_r$, decreases. 
Another characteristic feature, not observed frequently, 
is the absence of saturation of the magnetization for high external fields.
This phenomenon was already found in measurements of powder samples
of LCMO nanotubes, and was attributed to an antiferromagnetic behavior 
of the dead layer that surrounds the core of the nanoparticles \cite{Curiale2009}.
Later we will analyze this topic further.      

\section{Micromagnetic simulations \label{Simulations}}

A goal of this research is to use micromagnetic simulations 
to analyze in depth the experimental data. 
In this section, first we present our numerical scheme and then, 
it is used to calculate the hysteresis loops of two simple magnetic models. 
Once the best model to describe the behavior of the LCMO nanotubes has been chosen,
in the next sections we will take advantage of the numerical simulations 
to extract information about the characteristics of this system.        

\subsection{Micromagnetic simulation scheme}

As mentioned earlier, the LCMO nanotubes are composed of nanograins 
whose typical diameters range between $10$ nm and $40$ nm, 
which are smaller than the critical size of a single magnetic domain 
at $T=0$ K \cite{Curiale2004,Curiale2007}.
Also, since these nanoparticles are formed by thousands of atoms, 
at least at low temperatures the magnetization of each of them 
can be represented by classical vectors of magnitude equal to the saturation magnetization. 

In general, we describe the dynamic time evolution of such systems of classical magnetic nanoparticles 	
by the sLLG equation introduced by Brown \cite{Brown1963}
which, in the Landau formulation of dissipation \cite{Landau1935}, reads
\begin{eqnarray}
\frac{d\mathbf{M}_i}{dt}&=&-\frac{\gamma_0}{1+\eta_0^2} \mathbf{M}_i \nonumber \\
&& \times \left[\mathbf{H}_i+\mathbf{W}_i+ \frac{\eta_0}{M_s} \mathbf{M}_i
\times (\mathbf{H}_i+\mathbf{W}_i)\right],
\label{sLLG}
\end{eqnarray}
where $t$ is the time (in seconds), $\gamma_0\equiv\gamma \mu_0=2.2128 \times 10^5$ m/(As),
with $\gamma$ being the gyromagnetic ratio and $\mu_0$ the vacuum permeability constant,
and $\eta_0$ is an adimensional phenomenological damping constant.
$\mathbf{M}_i$ is the magnetization of the $i$th nanoparticle
whose magnitude is $M_s$, the saturation magnetization.
$\mathbf{H}_i$ is the local effective field acting at each site and is given by
\begin{equation} 
\mathbf{H}_i=-\frac{1}{\mu_0} \frac{\partial U}{\partial \mathbf{M}_i},
\label{EffectiveField}
\end{equation} 
where $U$ is the energy per unit volume of the system.
Thermal effects are introduced by random fields $\mathbf{W}_i$  
which are assumed to be Gaussian distributed with average 
\begin{equation} 
\langle W_{i,k}(t) \rangle_{\mathbf W} = 0 
\label{average}
\end{equation}
and correlations \cite{Brown1963}
\begin{equation} 
\langle W_{i,k}(t) W_{i,l}(t')  \rangle_{\mathbf W} = 2 D \ \delta_{kl} \ \delta(t-t'),
\label{correlations}
\end{equation}
for all $k,l=x,y,z$ components. The parameter $D$ is chosen as
\begin{equation}
D = \frac{\eta_0 k_B T}{\mu_0 \gamma_0 M_s V}, 
\label{paramaterD}
\end{equation}
so the sLLG equation takes the magnetization to equilibrium at temperature $T$.
Here, $k_B$ is the Boltzmann's constant and $V$ is the volume of each nanoparticle. 

The sLLG is a first-order stochastic differential equation
with a multiplicative thermal white noise coupled to magnetization. 
To preserve the magnetization module of each nanoparticle,
it is required to use the Stratonovich mid-point prescription \cite{Aron2014,Gardiner1997}.
The sLLG Eq.~(\ref{sLLG}) can be easily integrated using the Heun method which
converges to the solution interpreted in the sense of this 
explicit discretization scheme \cite{Rumelin1982,GarciaPalacios1998}.
In Cartesian coordinates, this integration method requires 
the explicit normalization of magnetization after every time step $\Delta t$ \cite{Martinez2004,Cimrak2007}.
We use a constant adimensional time step of $\Delta \tau = \gamma_0 M_s \Delta t = 0.01$, 
which is sufficiently small to ensure convergence from further reductions in $\Delta \tau$.
In addition, the simulations were performed choosing $\eta_0 = 0.01$.

\subsection{One-dimensional model}

Given the high aspect ratio of a nanotube, it is natural to assume that 
a one-dimensional model is enough to correctly describe its magnetic behavior.
This was the approach chosen recently in Ref.~\cite{Longone2018}, 
where micromagnetic calculations were performed to simulate 
a long chain of dipolar-interacting anisotropic single-domain particles.  
Using an appropriate set of parameters, with this simple model it was possible 
to fit well the hysteresis loop of single LCMO nanotubes measured at $T=14$ K.
Following the lines of Ref.~\cite{Longone2018}, here we have carried out 
micromagnetic calculations of this one-dimensional model 
to analyze our present experimental data, 
which have been measured for a very wide range of temperatures.  

The dynamics of the model is not very sensitive to the choice of the volume of nanograins, 
and therefore we consider that they all have the same diameter, $d$.
Structurally, the system consists of $L$ nanograins equally spaced along a linear chain,
with a separation $d$ between them.
Due to the existence of the dead layer, only long-range dipolar interactions are considered.
Disorder is introduced into the model considering that each nanograin has a particular uniaxial anisotropy.
This originates in the fact that in LCMO nanotubes, the nanoparticles 
have a non-spherical morphology with aspect ratios large enough for the shape anisotropy 
to dominate over the crystalline one of the manganite compound \cite{Curiale2007}.

The energy per unit volume of this one-dimensional model is given by
\begin{eqnarray}
U&=& -\mu_0 \ \mathbf{H} \cdot \sum_{i=1}^L \mathbf{M}_i 
-\frac{1}{M_s^2}\sum_{i=1}^L K_i (\mathbf{M}_i \cdot  \mathbf{n}_i)^2  \nonumber \\
&& -\frac{\mu_0 V}{4 \pi} \sum_{i<j} \bigg[ \frac{ 3(\mathbf{M}_i \cdot \mathbf{e}_{ij})
(\mathbf{M}_j \cdot \mathbf{e}_{ij})- \mathbf{M}_i \cdot \mathbf{M}_j}{d_{ij}^3} \bigg]. 
\label{energy}
\end{eqnarray}
The first term is the Zeeman interaction,
the second represents the anisotropy energy,
and the last is the dipolar coupling between nanoparticles. 
$V=\pi d^3/6$ is the volume of each nanograin and 
$L$ is the number of nanograins which are equally spaced and aligned along the $z$ axis.
$\mathbf{n}_i$ is the uniaxial anisotropy axis vector at site $i$ 
that is randomly oriented and $K_i$ the corresponding constant,
$\mathbf{e}_{ij}$ is a unit vector pointing from the site $i$ to the site $j$,
and $d_{ij}$ is the distance that separates these two points which is an integer multiple of $d$.
As before, $\mathbf{H}$ is the external magnetic field.

The shape anisotropy constant of each nanoparticle can be written as 
\begin{equation}
K_i=\frac{1}{2}\mu_0 N_i M_s^2 ,
\label{aniconst}
\end{equation}
where, for a prolate ellipsoid, $N_i$ is the difference between the demagnetizing 
coefficients along their major and minor axes \cite{Cullity}.
$N_i=0$ for a spherical body and $N_i=1/2$ for an infinitely long rod.
Note that, assuming that the shape of a nanograin (more precisely, the shape of the core) 
does not change with temperature, then $K_i$ depends on $T$ through $M_s$.
Although originally it was assumed a uniform distribution for the anisotropy constant,
motivated by a recent study on a similar magnetic system \cite{McGhie2017}, 
here we consider that $N_i$ is Gaussian distributed with 
mean $N_0$ and standard deviation $\sigma_N$. 
  
We simulate the dynamic of the model using the numerical scheme presented above.
From Eqs.(\ref{EffectiveField}) and (\ref{energy}), the effective field is
\begin{eqnarray}
\mathbf{H}_i &=& \mathbf{H} + \frac{2 K_i}{M_s^2 \mu_0} (\mathbf{M}_i \cdot  \mathbf{n}_i)  \mathbf{n}_i \nonumber \\
&& +\frac{V}{4 \pi} \sum_{j \ne i} \bigg[ \frac{ 3(\mathbf{M}_j \cdot \mathbf{e}_{ij}) \mathbf{e}_{ij}
- \mathbf{M}_j}{d_{ij}^3} \bigg]. 
\label{EffectiveField2}
\end{eqnarray}
The sum in Eq.~(\ref{EffectiveField2}), which is the contribution of the long-range dipolar interactions, 
extends to all sites except the $i$th one. 
The performance of our algorithm strongly depends on the strategy chosen to calculate this term.
Instead of using a sophisticated technique, like the Ewald \cite{LandauBinder2005} one, 
here we use the simpler Lorentz-cavity method \cite{Berkov2001}.
The idea is to perform explicitly the sum in Eq.~(\ref{EffectiveField2}) 
over the sites $j$ surrounding the $i$th site 
up to a certain lattice distance $l_c$ ($|i-j|\le l_c$) and, 
to calculate the contribution of the remaining terms, 
the corresponding values of magnetization
are considered to be equal to the mean value of magnetization of the system
\begin{equation}
\mathbf{\overline{M}}=\frac{1}{L}  \sum_{i=1}^L \mathbf{M}_i.
\end{equation}  
In other words, the least relevant part of the dipolar field 
(the one produced by the nanograins that are far from of site $i$)
is calculated using a kind of mean field approximation.
Although this method is not theoretically rigorous, 
we have verified that choosing $l_c=2$, we obtain curves
that differ from those calculated without using any approximation (whose runtime is huge)
by an amount that is less than the statistical errors.       

In a typical run, we average the mean value of magnetization,
\begin{equation}
\mathbf{M}=\langle \mathbf{\overline{M}} \rangle,
\end{equation}     
at different temperatures and external fields. 
$\langle ...\rangle$ represents an average over at least $10^2$ disorder realizations.  
In all cases, we simulate systems of size $L=10^2$ with $d=25$ nm, 
applying the external field along the $z$ axis. 
We calculate the hysteresis loops starting from a random magnetization state
and we sweep the external field at a given rate $R$.
Since integrating the sLLG equation requires to use very short-time steps,
for our computing capabilities it is possible 
to calculate only high-frequency hysteresis loops.  
Nevertheless, it has been shown that these quickly converge 
to a limit curve as $R$ decreases and,
therefore, it should not be very different from that 
obtained under experimental conditions \cite{Longone2018}.  
We have verified that using a rate of $R=10^{10}$ A/(m s) 
is enough to fulfill this requirement for temperatures up to $T=80$ K. 

Performing several micromagnetic simulations of the one-dimensional model, 
we have searched for the best set of parameters, 
$M_s$, $N_0$, and $\sigma_N$, that allow us to fit the experimental data at $T=10$ K. 
Note that with this model we are only able to describe 
the magnetic behavior of the ferromagnetic cores of the nanograins,
but not the contribution of their antiferromagnetic shells.
For this reason, we focus on describing the experimental data 
for low values of the external field, more precisely for $|H| \lesssim H_c$.

\begin{figure}[t!]
\begin{center}
\includegraphics[width=6.5cm,clip=true]{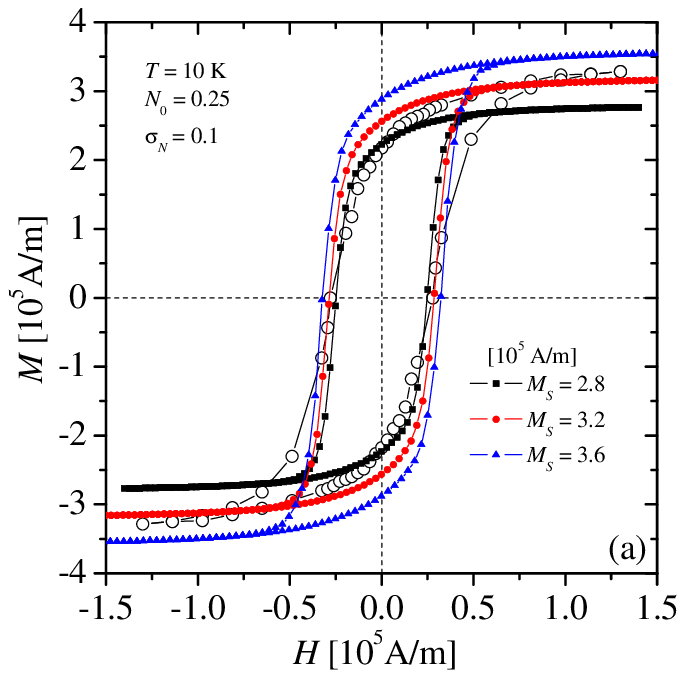}
\includegraphics[width=6.5cm,clip=true]{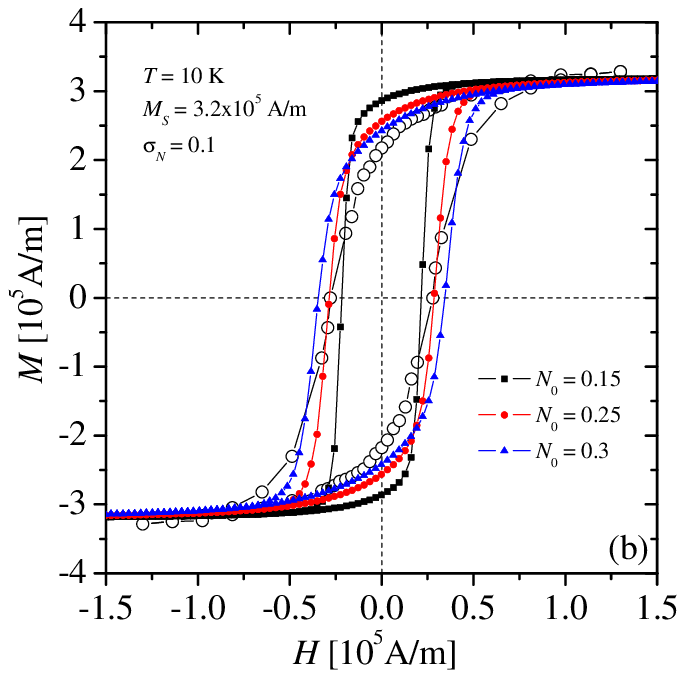}
\includegraphics[width=6.5cm,clip=true]{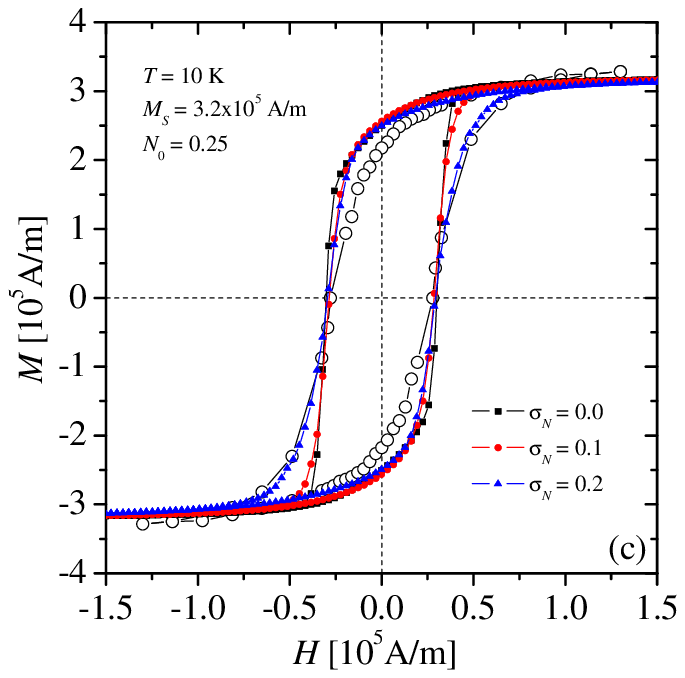}
\caption{(Color online) Comparison between the experimental data at $T=10$ K (black open circles),
and the hysteresis loops for the one-dimensional model calculated for three values of 
(a) $M_s$, (b) $N_0$, and (c) $\sigma_N$, 
keeping constant the corresponding remaining parameters as indicated.} 
\label{figure3}
\end{center}
\end{figure}

Figure~\ref{figure3} (a) shows the hysteresis loops at $T=10$ K
calculated numerically for three values of $M_s$,
keeping constant $N_0=0.25$ and $\sigma_N=0.1$ (for simplicity, we omit to plot the virgin curve).  
Simulation curves are compared with the experimental data obtained at this same temperature.  
We see that changes in the saturation magnetization produce variations in both,
the coercive field and the remanent magnetization, 
without greatly affecting the characteristic shape of the curves.
This behavior can be explained as follows.
When increasing $M_s$ the hysteresis loops, for high external fields, 
must tend to a higher asymptotic magnetization values, 
which produces a vertical expansion of the curves and therefore an increase in $M_r$.
Also, from Eq.~(\ref{aniconst}), this increase in $M_s$ is accompanied by an increment 
of the anisotropy constant of each nanoparticle and, consequently,
$H_c$ has to increase as well.

A qualitatively different behavior is observed if we take constants  
$M_s=3.2 \times 10^5$ A/m and $\sigma_N=0.1$,
and we vary the parameter $N_0$, Fig.~\ref{figure3} (b).
Now the curves tilt and widen as $N_0$ increases and,
although the coercive field again increases (since the anisotropy constant $K_i$ increases), 
in this case the remanent magnetization tends to decrease.
This last feature can be explained in very simple terms.
The increase in $N_0$ makes the contribution of the anisotropy 
to the effective field Eq.~(\ref{EffectiveField2}),
which tends to align the magnetic moments in the direction 
of the randomly oriented axes $\mathbf{n}_i$ thus decreasing the value of $M_r$,
exceeds that of the dipolar one (since we have kept $M_s$ constant), 
whose tendency is to keep the moments aligned, thereby increasing $M_r$. 
The net effect is a reduction of the remanent magnetization.

Implementing one of these two strategies is not possible
to achieve a good fit of the experimental data. 
Fortunately, if we vary the parameter $\sigma_N$ 
(which controls the width of the distribution of $N_i$),
taking constant $M_s=3.2 \times 10^5$ A/m and $N_0=0.25$,
we can tilt the hysteresis loops keeping approximately invariant their coercive fields. 
Figure~\ref{figure3} (c) shows that choosing $\sigma_N=0.2$ (blue closed triangles)
it is possible to obtain a reasonable fit within the first and third quadrants,
although the simulation curve clearly does not describe the magnetization reversal process well. 
This is the best set of parameters we have been able to identify 
which allow us, at least partially, to meet our goal of describing experimental data accurately.
Nevertheless, for higher temperatures, the discrepancies 
between the experimental and simulation hysteresis loops
are more pronounced.  
  
\subsection{Model with dipolar lateral interactions}

The differences between the experimental and simulation data 
are mainly observed in the second and fourth quadrants of Fig.~\ref{figure3} (c).
This phenomenon may be caused by two kinds of mechanisms.
On the one hand, at finite temperature, 
it would be possible that the magnetization of nanoparticles 
breaks into a multidomain structure,
which would explain why $M_r$ has such a low value and 
why the reversal magnetization process is more efficient than in the simulation.
In that case, our one-dimensional model could never correctly describe this system. 

On the other hand, assuming the nanoparticles are still monodomain,
it is possible to recreate in the simulations the same behavior observed in the experiments, 
including in the effective field Eq.~(\ref{EffectiveField2}) the contribution of the nanograins 
surrounding those that lie along the linear chain
(that were already implicitly excluded when we defined the one-dimensional model).
For example, in a configuration for which $M>0$ and $H>0$, second quadrant,
the dipolar field along the chain produced by these nanoparticles 
that are located laterally off the $z$ axis,
should point in the same (opposite) direction to the applied magnetic field (magnetization).
Therefore, the overall effect due to the inclusion of these nanograins
will be to increase the effective field, decreasing further the magnetization of the system
during the reversion process.  
   
\begin{figure}[t!]
\begin{center}
\includegraphics[width=6.5cm,clip=true]{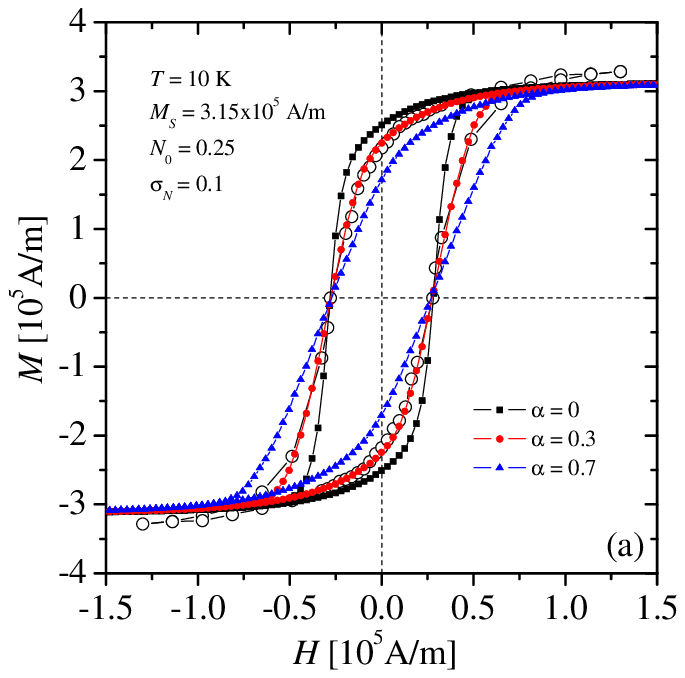}
\includegraphics[width=6.5cm,clip=true]{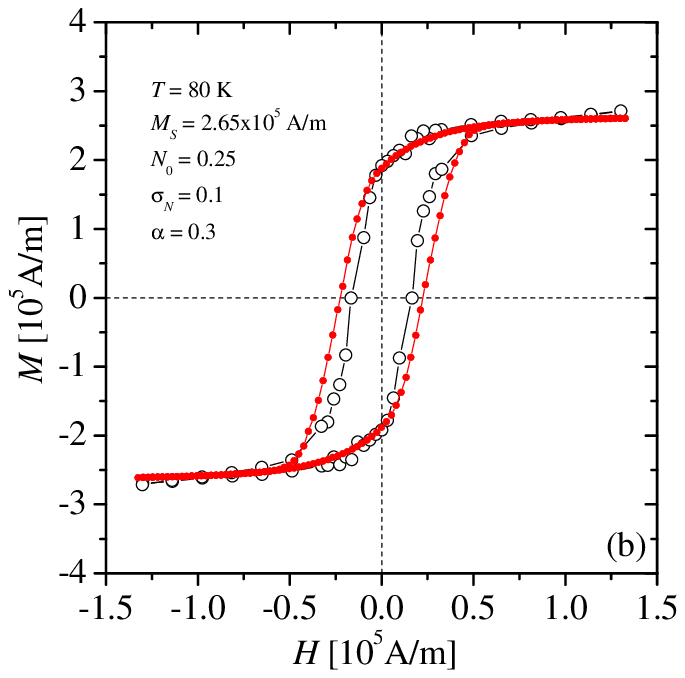}
\caption{(Color online) Comparison between the experimental data (black open circles)
and the hysteresis loops for the model with lateral interactions,
at (a) $T=10$ K and (b) $T=80$ K.
Simulation curves (closed symbols) were calculated for different
sets of parameters as indicated.} 
\label{figure4}
\end{center}
\end{figure}

\begin{figure}[t!]
\begin{center}
\includegraphics[width=6.5cm,clip=true]{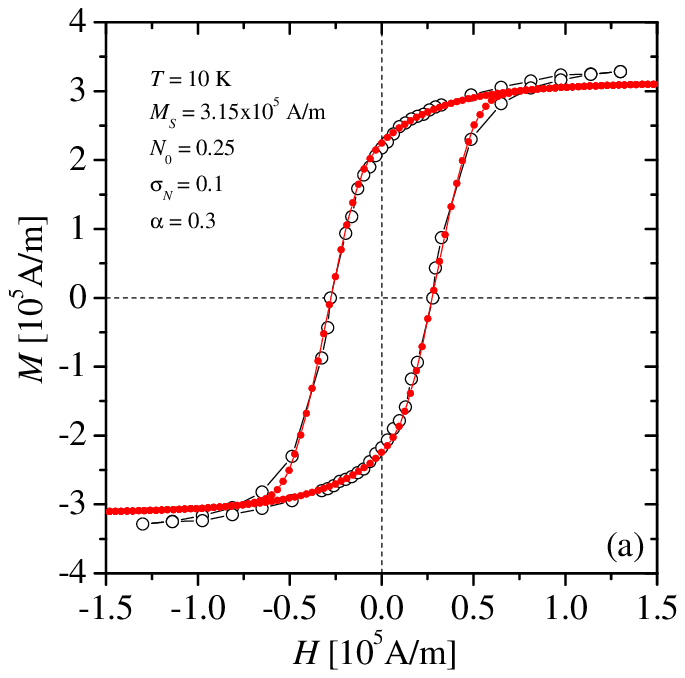}
\includegraphics[width=6.5cm,clip=true]{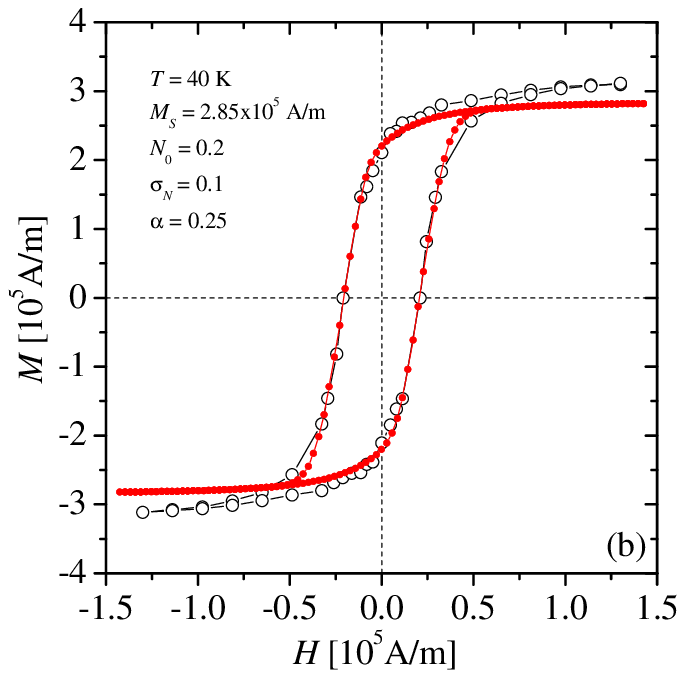}
\includegraphics[width=6.5cm,clip=true]{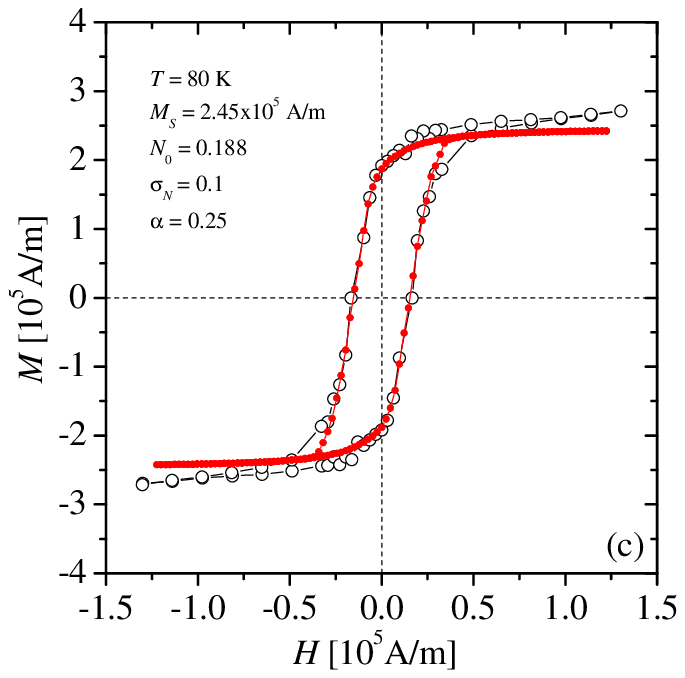}
\caption{(Color online) Idem to Fig.~\ref{figure4} but for
(a) $T=10$ K, (b) $T=40$ K, and (b) $T=80$ K,
and another set of parameters as indicated.  } 
\label{figure5}
\end{center}
\end{figure}

We modify our model to implement this last approach.
For simplicity, we assume that the nanograin at site $i$
feels a dipolar field produced by its surroundings, which we represent by
four magnetic moments located laterally at a distance $d$,
whose magnetizations are $\alpha \mathbf{\overline{M}}$.
$\alpha$ is a parameter that allows us to set the magnitude of this lateral interaction.
We add to the effective field Eq.~(\ref{EffectiveField2}) the term
\begin{equation}
\mathbf{H}_i^\textrm{lat} = -\frac{V \alpha }{4 \pi d^3} 
\sum_{j=1}^4 [3 (\mathbf{\overline{M}} \cdot \mathbf{e}_j)\mathbf{e}_j- \mathbf{\overline{M}}], 
\label{lateral_EffectiveField}
\end{equation} 
where the sum on $j$ is over the four magnetic moments 
located at positions $\mathbf{e}_1=\mathbf{e}_x$, $\mathbf{e}_2=\mathbf{e}_y$, 
$\mathbf{e}_3=-\mathbf{e}_x$, and $\mathbf{e}_4=-\mathbf{e}_y$,
with $\mathbf{e}_x$ and $\mathbf{e}_y$ being unit vectors pointing 
along the $x$ and $y$ axes, respectively. 
Note that we introduce lateral interactions in the form of a mean-field approximation,
with an environment that does not break the axial symmetry of the system.

We return to the basic problem of fitting the experimental data at $T=10$ K.
Figure~\ref{figure4} (a) shows the simulation curves for three values of $\alpha$
keeping constant $M_s$, $N_0$, and $\sigma_N$.
The behavior of these hysteresis loops is similar to one observed
in Fig.~\ref{figure3} (c) for the model without lateral interactions, 
when we vary the parameter $\sigma_N$.    
But now the fit obtained for $\alpha=0.3$ is much better than that previously achieved.
The reason is that in the reversal process, 
the new term that is added to the effective field, Eq.~(\ref{lateral_EffectiveField}), 
contributes to further decrease the value of magnetization.

We can interpret this result as follows.
Because the manganite nanotubes have a very disordered and irregular granular structure, 
and in our model lateral interactions have been included through a mean field approximation which, 
ultimately, is quite rudimentary (but very effective),
it is not possible to make a rigorous interpretation of the value of $\alpha$ obtained by our fit. 
However, if each nanograin were surrounded by four neighbors located at a distance $d$ 
then, ideally, $\alpha$ should be close to $1$.
We obtain $\alpha=0.3$, indicating that the lateral interactions or, in other terms, 
the demagnetizing field produced by the surrounding medium, 
it is of less intensity than in this ideal situation.
This is probably due to the nanotube wall being very thin:
For a tubular geometry with a radius ratio $s_r=a'/a$
($a'$ and $a$ being the inner and outer radii, respectively),
it is known that the demagnetizing factor decreases appreciably when $s_r$ tends to one 
(in our case $s_r \gtrsim 0.9$)\cite{Kobayashi1996}.    

Using the same set of parameters $N_0=0.25$, $\sigma_N=0.1$, and $\alpha=0.3$,
and a suitable value for the saturation magnetization,
it should be possible to fit the experimental data at any other temperature. 
Figure~\ref{figure4} (b) shows the result obtained at $T=80$ K following this strategy.
Although the value of the remanent magnetization calculated in the simulation 
matches very well with that obtained experimentally, 
the same does not hold true for the coercive field.

However, we obtain a better agreement with the experimental data at all temperatures
if we reconsider the initial hypotheses about the temperature independence of  
the anisotropy constant Eq.~(\ref{aniconst}) and the parameter $\alpha$. 
In addition to $M_s$, $N_i$ would not be anymore constant
if the aspect ratio of the core of the nanoparticles changes with $T$.     
This same effect should also affect the magnitude of the lateral interactions, 
i. e., the value of $\alpha$.

Figures~\ref{figure5} (a), (b), and (c), show the hysteresis loops
calculated, respectively, at $T=10$, $40$, and $80$ K,
allowing changes in the parameters $N_0$ (keeping constant $\sigma_N=0.1$) and $\alpha$. 
In this way, we achieve to fit very well all the experimental data available for single LCMO nanotubes.  
Table \ref{table} presents the parameters that we have used 
for each temperature between $T=5$ and $80$ K.

\begin{table}[t]
\begin{center}
\caption{\label{table} Simulations parameters $M_s$, $N_0$, $\sigma_N$, and $\alpha$,
as well the aspect ratio $r$, for seven temperatures up to $T=80$ K.}
\begin{tabular}{cccccc}
\hline
\hline
$T$(K)& \ $M_s$(A/m)				& \ $N_0$	& \ $\sigma_N$	& \ $\alpha$  & \ $r$  		\\ 
\hline
$5$	& \ $3.20 \times 10^5$	& \ $0.270$	& \ $0.1$		& \ $0.30$	  & \	$2.24$	\\
$10$	& \ $3.15 \times 10^5$	& \ $0.250$	& \ $0.1$		& \ $0.30$	  & \	$2.08$	\\
$15$	& \ $3.10 \times 10^5$	& \ $0.240$	& \ $0.1$		& \ $0.30$	  & \ $2.01$	\\
$20$	& \ $3.05 \times 10^5$	& \ $0.230$	& \ $0.1$		& \ $0.30$	  & \ $1.93$   \\
$40$	& \ $2.85 \times 10^5$	& \ $0.200$	& \ $0.1$		& \ $0.25$	  & \	$1.75$   \\
$60$	& \ $2.65 \times 10^5$	& \ $0.190$	& \ $0.1$		& \ $0.25$	  & \ $1.69$   \\
$80$	& \ $2.45 \times 10^5$	& \ $0.188$	& \ $0.1$		& \ $0.25$	  & \ $1.68$   \\
\hline
\hline
\end{tabular}
\end{center}
\end{table}

\section{Results and conclusions \label{ResCon}}

So far, we have shown how using a micromechanical torsional oscillator, 
the hysteresis loops of two LCMO nanotubes of manganite 
could be measured at different temperatures between $T=5$ and $80$ K.
Then, performing extensive micromagnetic simulations, 
it was possible to determine that it is necessary 
to go beyond a one-dimensional model of nanotubes 
to fit the experimental data well.
In fact, it was essential to include dipolar lateral interactions 
to correctly describe the process of magnetic reversal.  
Thus, we have reached the ability to simulate 
the ferromagnetic response of these nanostructures
under the influence of a longitudinal external field. 

Now, we use the numerical data to study more in depth these manganite nanostrutures.
From the comparison between the experimental and simulation hysteresis loops
shown in Figs.~\ref{figure5} (a), (b), and (c),
we can see how much the dead layer influences the magnetic behavior of nanotubes.
For external fields of magnitude greater than the coercive one, 
the magnetization begins to deviate away from the value $M_s$ approximately in a linear way.
This same behavior was observed in powder samples
of LCMO nanotubes, which was interpreted as evidence of the 
antiferromagnetic character of the dead layer \cite{Curiale2009}.
As our measurements have been made on individual nanostructures then, 
in addition to confirming the existence of this phenomenon, 
we can rule out that it is caused by other factors 
that are present in powder samples. 
Namely, by the interactions between nanotubes or simply 
by their orientations that are distributed isotropically at random. 

\begin{figure}[t!]
\begin{center}
\includegraphics[width=6.5cm,clip=true]{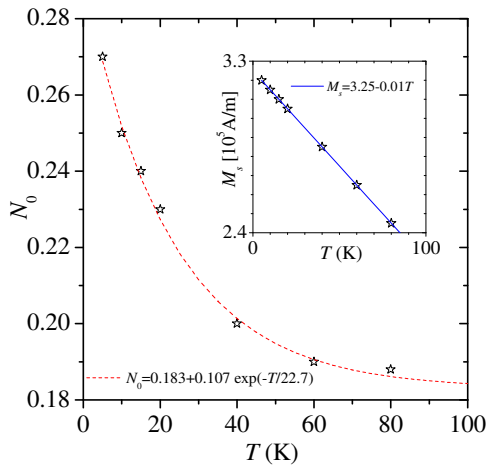}
\caption{(Color online) Plot of the parameter $N_0$ used in the simulations (open stars)
as function of temperature up to $T=80$ K.  Data has been taken from Table \ref{table}. 
The dashed red line corresponds to Eq.~(\ref{N_0}).
Inset: The same but for $M_s$.  The continuous blue line correspond to Eq.~(\ref{M_s}).} 
\label{figure6}
\end{center}
\end{figure}

To achieve good fits of the experimental data, 
it was necessary to use the parameters given in Table \ref{table}.
Figure~\ref{figure6} shows a plot of $N_0$ versus $T$.  
The temperature dependence of this quantity
can be well described by the exponential function,
\begin{equation}
N_0 = 0.183+0.107 \exp(-T/22.7).
\label{N_0}
\end{equation}
Instead, $M_s$ follows a linear law,
\begin{equation}
M_s = 3.25-0.01 T,
\label{M_s}
\end{equation}
see inset in Fig.~\ref{figure6}.
A similar behavior of the saturation magnetization with $T$
was previously measured in this same system \cite{Dolz2008}.
Nevertheless, we emphasize that the approach we have used here
allows us to calculate the contribution of the core to $M_s$, 
while a direct analysis of the experimental data (as done in Ref.~\cite{Dolz2008}) 
is not enough to eliminate the contribution of the dead layer to $M_s$.

\begin{figure}[t!]
\begin{center}
\includegraphics[width=6.7cm,clip=true]{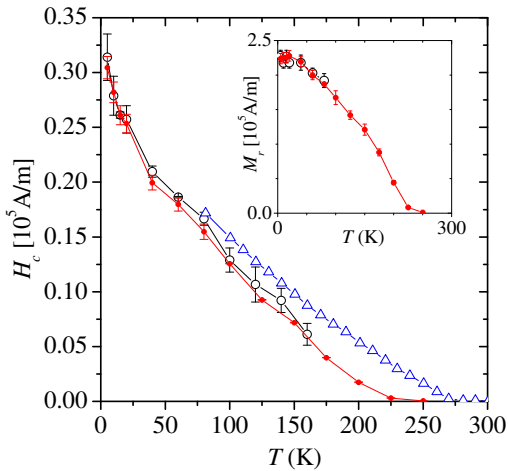}
\caption{(Color online) Coercive field as function of $T$ 
measured for two single LCMO nanotubes (black open circles),
for a powder sample (blue open triangles)  \cite{Curiale2007}, 
and calculated through micromagnetic simulations (red closed circles).
Inset shows the temperature dependence of remanent magnetization.} 
\label{figure7}
\end{center}
\end{figure}

Taking constants $\alpha=0.25$ and $\sigma_N=0.1$, we use Eqs.(\ref{N_0}) and (\ref{M_s})
to perform new micromagnetic simulations up to $T=250$ K.  
For $T \ge 100$ K, we calculate the hysteresis loops using a rate of $R=10^{9}$ A/(m s). 
Figure~\ref{figure7} shows the coercive field versus the temperature
measured for the two single LCMO nanotubes, and also computed through our simulations.
Although, due to experimental limitations in the sensitivity and efficiency
of our micromagnetometer, we could only measure 
the complete hysteresis loops up to $T=80$ K,
for higher temperatures, we were able to estimate $H_c$
performing several cycles averaging the inverse field values 
that were necessary to apply to reach the condition $\Delta \nu \approx 0$.

On the one hand, we observe in Fig.~\ref{figure7} that there is a very good agreement 
between our calculated value of $H_c$ and the one experimentally measured up to $T=160$ K.
This result suggests that our model, and more precisely the assumption 
that the nanograins behave like the single magnetic domain,
could be valid over a broad range of temperatures. 
On the other hand, at $T \approx 250$ K the simulations 
indicate that $H_c$ falls to zero. 
In contrast, the coercive field measure for a powder sample (see Fig.~\ref{figure7})
tends to zero at $T \approx 273$ K \cite{Curiale2007},
a value that perfectly matches the critical temperature 
in the bulk, $T_c \cong 273$ K \cite{Cheong2004}. 
This suggests that some of the properties measured for powder samples,
could depend to a greater extent on the type of magnetic material being studied, 
than on its structure at the nanometric level.

In the inset of Fig.~\ref{figure7}, we show the values
of remanent magnetization measured for the two single LCMO nanotubes 
up to $T=80$ K, and computed using the micromagnetic calculations.
As before, we find that there is a good coincidence between both data sets and, 
from the simulations, we again observe that at $T \approx 250$ K
hysteresis disappears and therefore $M_r$ falls to zero.

\begin{figure}[t!]
\begin{center}
\includegraphics[width=6.4cm,clip=true]{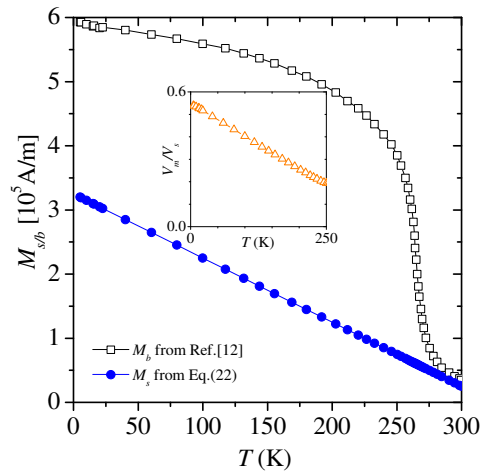}
\caption{(Color online) The saturation magnetization versus
the temperature given by Eq.~(\ref{M_s}) (blue closed circles)
and for a LCMO bulk sample \cite{Dolz2008}.
Inset shows the fraction $V_m/V_s$ as function of $T$, see text.} 
\label{figure8}
\end{center}
\end{figure}

The simulation parameters given in Table \ref{table},
and in particular the extrapolations for higher temperatures,
Eqs.(\ref{N_0}) and (\ref{M_s}), 
unveil that the core undergoes significant changes of volume 
but more subtle in shape.
Figure~\ref{figure8} shows a comparison between 
the saturation magnetization given by Eq.~(\ref{M_s}), 
and the measure for a LCMO bulk sample, $M_b$ \cite{Dolz2008}.
We observe that there are important differences
both in their temperature dependencies 
and in their magnitudes.
The origin of this phenomenon is a highly discussed issue that has not yet been clarified \cite{Caizer2015}.
Nevertheless, a possible physical picture of the mechanisms involved is given below.

{\em A priori}, there should be no differences between 
the saturation magnetizations of both, the core (which we assume is uniformly magnetized),
calculated as the ratio between the magnitude of 
its magnetic moment and its volume, and the bulk sample. 
However, in our experimental measures of single LCMO nanotubes
and also in our micromagnetic simulations, 
we have calculated the magnetization using geometric volumes (of nanotubes or nanograins). 
Therefore, the discrepancies between $M_b$ and $M_s$ observed in Fig.~\ref{figure8},
could be interpreted as evidence of changes with temperature in the volume of the core \cite{Caizer2003}.

To quantify this change, we consider that $M_s=m/V_s$, 
where $m$ is the magnetic moment of the sample and $V_s$ its geometric volume.
Instead, the saturation magnetization of the core, 
which we consider to be equal to that of bulk, is $M_b=m/V_m$, 
$V_m$ being the effective magnetic volume of nanotubes (the total volume occupied by the cores).  
So, the fraction of the sample that is ferromagnetic is 
\begin{equation}
\frac{V_m}{V_s}=\frac{M_s}{M_b}.
\end{equation}
In the inset of Fig.~\ref{figure8}, we show the dependence on temperature of this ratio.
While at $T=0$ K $V_m/V_s \sim 0.54$, at $T=250$ this ratio falls sharply to approximately $0.2$.

To explain the change of the effective magnetic volume with temperature, 
we use the model proposed in Ref.~\cite{Caizer2003}
originally intended to describe this phenomenon in
$\gamma$-Fe$_2$O$_3$ ferrimagnetic nanoparticles dispersed in a silica matrix. 
Most likely the disordered surface structure of the manganite nanograins (dead layer)
leads to a weakening of the double exchange interactions $J$ between Mn ions \cite{Curiale2009}.
Since this distortion decreases progressively toward the core,
we assume that the structure of a nanoparticle is made up of several sub-layers $i$, 
each of which is characterized by a $J_i$.
This exchange interaction increases gradually from a minimum value at the surface
to a maximum one at the center of the core. 
As the critical temperature of each sub-layer $T_{c,i}$ is roughly proportional to $J_i$,
at temperature $T$ only the shells with $T_{c,i}>T$ will be ordered ferromagnetically.
Therefore, the magnetic volume of the nanoparticles will increase as the temperature decreases.   
This effect would explain the changes in the saturation magnetization 
of manganite LCMO nanotubes observed in this work. 

In addition, the dependence on temperature of $N_0$
suggests that the ferromagnetic core
of the nanoparticles undergoes slight shape changes as $T$ increases. 
Table \ref{table} indicates for each value of $N_0$, 
the corresponding aspect ratio $r$ that a typical core should have 
if we assume that the shape of this resembles that of
a prolate ellipsoid \cite{Cullity}.
Our findings indicate that $r$ tends to decrease as the temperature increases.
In a sense, this result is not surprising because as the fraction $V_m/V_s$ decreases,
the influence of the surface of the grain in the core should also decrease and, 
therefore, its shape should slowly tend (even if it does not reach it) to that of a sphere.

In conclusion, in this work we used a silicon micromechanical torsional oscillator 
to measure the hysteresis loops of two manganite LCMO nanotubes at different temperatures.
Micromagnetic calculations are performed first to validate a simple model that allows 
quantitatively describing the ferromagnetic behavior of the system,
and then to study the experimental data more in depth.
The temperature dependence of the coercive field and remanent magnetization 
indicate that the hysteresis ceases to exist at $T \approx 250$ K, 
a lower value than the corresponding one measured for powder samples which,
in turn, is equal to the critical temperature of the bulk manganite LCMO, $T_c \cong 273$ K.
In addition, from the dependence of $M_s$ and $N_0$ with $T$,
we deduce that the ferromagnetic cores of nanoparticles 
undergo significant changes in volume and shape with temperature. 
 
\begin{acknowledgments}
We would like to thank A.G. Leyva for providing the samples of manganite nanotubes.
This work was supported in part by CONICET under Project No. PIP 112-201301-00049-CO 
and by Universidad Nacional de San Luis under Project No. PROICO P-31216 (Argentina). 

\end{acknowledgments}

\end{document}